\documentclass[twocolumn,preprintnumbers,amsmath,amssymb,floatfix,superscriptaddress]{revtex4}

\usepackage{graphicx}
\usepackage{amsmath}
\usepackage{amssymb}
\usepackage{bm}
\usepackage{mathrsfs}
\usepackage{color}
\usepackage[normalem]{ulem}

\begin{document}

\title{Renyi entropy flows from quantum heat engines}
\author{Mohammad H. Ansari}
\author{Yuli V. Nazarov}
\affiliation{Kavli Institute of Nanoscience, Delft University of Technology, P.O. Box 5046, 2600GA Delft, The Netherlands}

\date{\today}
\begin{abstract}
We evaluate R\'{e}nyi entropy flows from generic quantum heat engines (QHE) to a weakly-coupled probe environment kept in thermal equilibrium. We show that the flows are determined not only by heat flow but also by a quantum coherent flow that can be separately measured in experiment apart from the heat flow measurement.  The same pertains to Shanon entropy flow.  This appeals for a revision of the concept of entropy flows in quantum nonequlibrium thermodynamics.
\end{abstract}

\pacs{05.30.-d; 03.67.-a; 03.67.Mn,42.50.Ct}

\maketitle

\section{Intoduction}

Entropy production in heat engines has been a key concept in establishing the fundamental laws of thermodynamics \cite{Mazur}.   Recently, the laws have been reconsidered  for small systems in the context of fluctuation relations \cite{relations}, which  gave rise to much of experimental \cite{experimental} and theoretical \cite{theoretical} research. It is worth noting that the fluctuation relations are traditionally formulated in terms of entropy production that is computed with the definition described in  Ref.  \cite{Seifert}. 
While the definition is perfect for classical states, its validity needs to be revisited in quantum mechanics, where Shanon entropy is nonlinear in density matrix and its change is not necessarily related to the expectation value of any operator, and therefore its measurability is questionable \cite{Camprisi}. 


A quantum heat engine is a system of several discrete quantum states and, similar to a common heat engine, is connected to several environments kept at different temperatures. The motivation for research in QHE comes from studying models of photocells and photosynthesis \cite{Scully}. It has been demonstrated that quantum effects can dramatically
change the thermodynamics of QHEs \cite{drastic} and their fluctuations \cite{Rahav} manifesting the role of quantum coherence. We need to stress that the mere presence of discrete quantum states in the engine is not enough to reveal the coherence. The effects come from non-diagonal elements of the engine density matrix that require a coherent drive
and/or degeneracy of the engine states to facilitate the formation of  quantum superpositions \cite{Rahav}.

A generalization of Shanon entropy is the R\'{e}nyi entropy \cite{renyi} defined
here  (see Appendix A) as $S_M \equiv {\rm Tr}[\hat{\rho}^M]$, with $\rho$ being the density matrix of a quantum system. Shanon entropy $S$ is obtained from $S_M$ by taking a formal limit $S= \lim_{M \to 1} \partial S_M/\partial M$. Much theoretical research addresses R\'{e}nyi entropies in strongly interacting systems {\cite{{r1},{lieb}}}, in particular spin chains \cite{Korepin}.  Since $S_M$ is not linear in density matrix, its observability is not evident: some tricks \cite{Ivanov} may help in certain situations.  Similar to the flows of physical conserved quantities, such as charge and energy, conserved measures, such as Shannon and Renyi entropies, flow between subsystems. 

Recently, one of the authors has proposed a method for consistent quantum evaluation of R\'{e}nyi entropy flows, R-flows \cite{Nazarov}, defined as $\mathcal{F}_M\equiv -d\ln S_M/dt=-(dS_M/dt)/S_M$. The Shanon entropy flow ${\cal F}_S$ is obtained by taking limit ${\cal F}_S =\lim_{M \to 1}  \partial {\cal F}_M/\partial M$. In this paper, we  adjust and apply this method for QHE.


We evaluate R-flows from a generic heat engine to a {\it probe} environment that is weakly coupled \footnote{The assumption of weak coupling, which allows us to stick to second order perturbations, is crucial for us. It can be violated in two ways: First, the higher-order connected diagrams may, as discussed in \cite{Nazarov},  break the correspondence between entropy and heat flows. Second, the second order diagrams of the probe environment can complete with those of other environments, we discuss this at the end of this article.} with the engine,  thus not disturbing its workings. We find that R-flows can be naturally separated into incoherent and coherent parts; this also pertains to the flow of Shanon entropy. The incoherent part is related to the heat flow $Q$ to the environment.  For Shanon entropy flow we recover the textbook formula ${\cal F}_S = Q/T$, $T$ being the temperature of the probing environment. The coherent part is specific for coherent drive and is proportional to the second power of the density matrix of the engine. This raises concerns about its observability. However, the coherent part can be accessed in an experiment that is different from heat flow measurement: There, one measures averaged forces acting on the environment and computes the would-be dissipation due to these forces. While this fictitious dissipation is \emph {not} the heat flow, it does contribute to the entropy flows.  

\begin{figure}[tb]
\centerline{\includegraphics[width=0.8\linewidth]{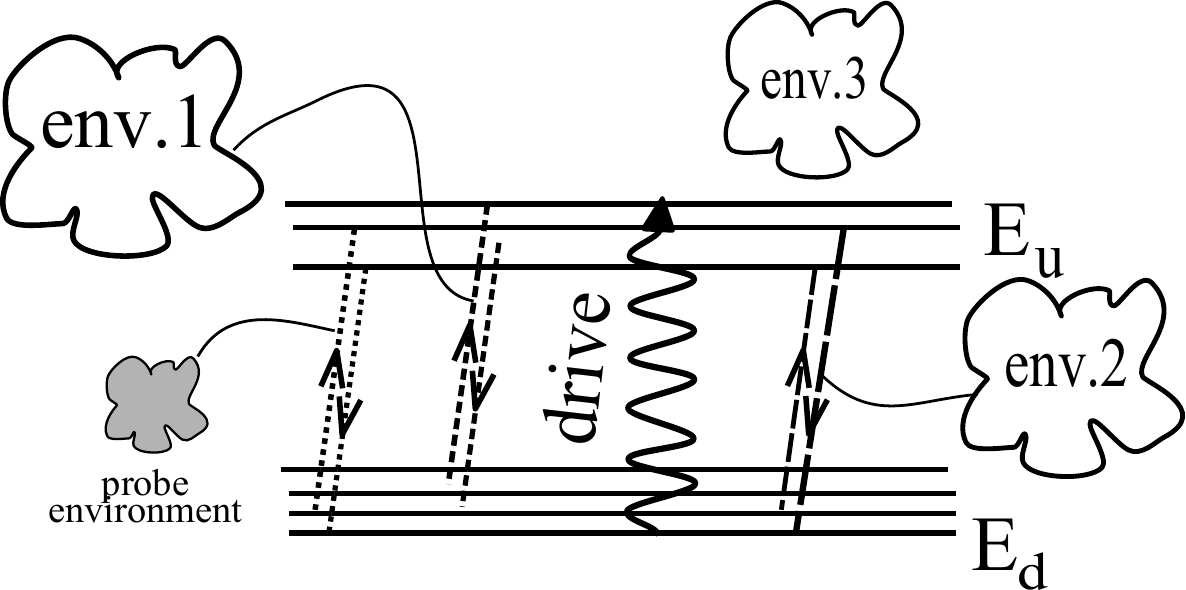}}
\caption{Schematics of QHE. A quantum system with two sets of states separated by energy $E_u -E_d$ is driven by external field at matching frequency. The system interacts with a number of environments inducing transitions between the states. We study the R-flows to a weakly coupled probe environment. }
\label{fig_engine}
\end{figure}

\section{General model: system and environments}

We consider a quantum system with discrete states $|n\rangle$  separated into two sets $\{u\},\{d\}$. All states within a set have approximately the same energy $E_{u}(E_{d})$,  the splitting $\epsilon_n$ within a set being much smaller than $E_u -E_d >0$. The system is subject to the external field with the frequency $\omega \approx E_u -E_d$ (we set $\hbar, k_B=1$ where appropriate) described by the Hamiltonian 
$H_{dr} = \sum_{m,n} \Omega_{mn} |m\rangle\langle n|  e^{-i\omega t} +H.c.$, and the relevant matrix elements are between the states of two sets. 

The quantum system is coupled to a number of environments labeled by $a$ kept at different  temperatures $T_a$.  The interaction with the environment is described by $H_{int}=  
\sum_{mn}|m\rangle\langle n|\hat{X}^{(a)}_{mn}$, with $\hat{X}^{(a)}_{mn}$
being the operators in the space of environment $a$. We assume linear response of each environment on the state of quantum system. In this case, each environment is completely characterized by the set of frequency-dependent generalized susceptibilities
$\chi^{(a)}_{mn,pq}(\nu)$ that are related to the correlators of {$\hat{X^{a}}$ defined as $S^{(a)}_{mn,pq}(t) \equiv \textup{Tr}_a \{\hat{X}^{a}_{mn}(0)\hat{X}^{a}_{pq}(t)\rho_a\}$}. The fluctuation-dissipation theorem yields the relations in frequency domain: {$S_{mn,pq}(\nu) =  n_B(\nu/T) \tilde{\chi}_{mn,pq}(\nu)$} where 
{$\tilde{\chi}_{mn,pq}(\nu) \equiv (\chi_{mn,pq}(\nu)-\chi_{pq,mn}(-\nu))/i$},
and  the Bose distribution ${n}_B(\nu/T) \equiv 1/ (\exp({\beta\nu})-1)$.
   
The environments produce  {transition} rates between the states of quantum system and affect the coherence of its density matrix $\rho_{nm}$. The dynamics are expressed by Bloch-master equation in the rotating wave approximation:   
\begin{eqnarray}
\label{eq:Bloch}
 &&\frac{d\rho_{mn}}{dt}= -i \sum_p(H_{mp}\rho_{pn} -\rho_{mp}H_{pn}) \\ &&  - \frac{1}{2} \sum_p(\Gamma_{mp}\rho_{pn} +\rho_{mp}\Gamma_{pn} )  +\sum_{p,{q}}\Gamma_{mn,pq} \rho_{pq} \nonumber 
\end{eqnarray}
To distinguish the sets, let us introduce a matrix $\eta_{nm}$, $\eta_{nm} = 1$ if $n \in \{u\}$ and $m \in \{d\}$, $\eta_{nm} = -1$ if $n \in \{d\}$ and $m \in \{u\}$,
$\eta_{nm}=0$ otherwise. The residual Hamliltonian is composed of three groups of terms
\begin{align}
H_{nm} = \epsilon_n \delta_{nm} + {\rm Re}\Omega_{nm}\eta^2_{nm} + i \eta_{nm} {\rm Im}\Omega_{nm}  \\+ \sum_{a,k} \int \frac{d\nu}{2\pi} \frac{S_{nk,km}(\nu)}{\nu - E_k+E_m} \nonumber
\end{align}
The first term is the original small splitting of the states, the second and third terms represent the coherent drive and the last term is the renormalization due to the interaction with the environments.  The dissipative terms $\Gamma$ are sums over the contributions of each environment, 
\begin{equation}
{\Gamma_{mn,pq} \equiv \sum_a S^{(a)}_{qn,mp}(E_{mp});\; \Gamma_{mn} \equiv \sum_{k}\Gamma_{{k}k,{m}n}}
\end{equation}
{ with $E_{mp}\equiv E_m-E_p$. The relevant terms satisfy $E_{mp}\approx E_{nq}$.} In the rotating wave approximation we can replace $E_{mn}$ with $\omega \eta_{mn}$.  In the absence of the drive and for the non-degenerate states the only relevant {$\Gamma$'s} are the transition rates from $m$ to $n$, the density matrix is diagonal and the equation reduces to the master equation. 

The Bloch equation (\ref{eq:Bloch}) can be obtained 
by time-dependent perturbation theory for density matrix in time interval {$(-\infty, t]$} where evolution operators for bra and ket are expanded in terms of $\hat{X}$, this sets the time ordering along the Keldysh contour that has opposite time directions for bra and ket (left diagram in Fig.\ref{fig_worlds}) For relevant diagrams the $\hat{X}^{(a)}$  are pairwise grouped, and the result of tracing over the environment is readily expressed in terms of {$S_{mn,pq}(t)$}. 
 The density matrix $\hat{\rho}(t)$ is obtained by summation over all such diagrams. The compact way to achieve the summation is to take the diagrams ending at $\tau=t$ and thus contributing to $d\hat{\rho}/dt$ at $\tau =t$ and replace $\hat{\rho}(-\infty)$ with $\hat{\rho}(t)$: This reproduces  Eq. (\ref{eq:Bloch}).

\section{Perturbation theory}

\begin{figure}[tb]
\centerline{\includegraphics[width=0.95\linewidth]{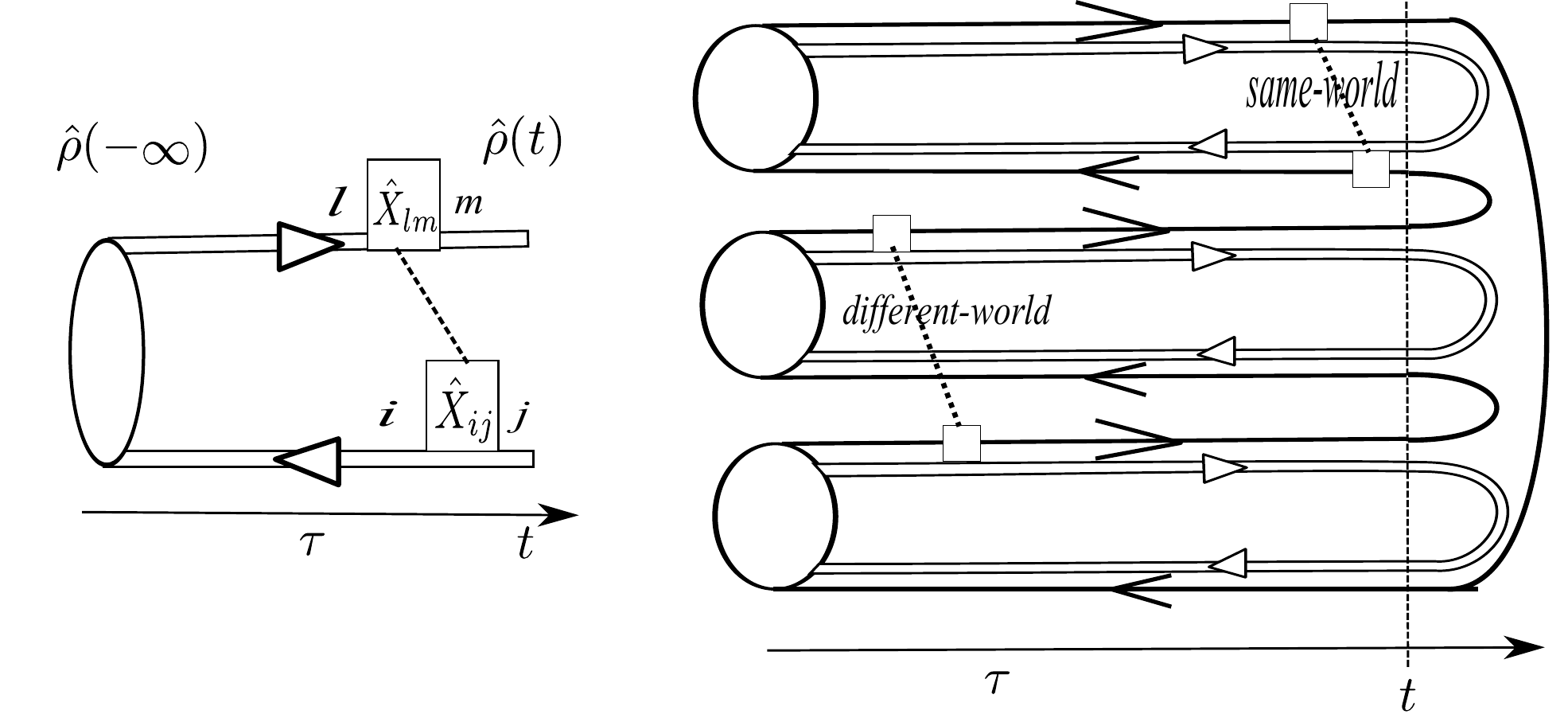}}
\caption{ Left: Perturbation  on the Keldysh contour for a single world, where operator $\hat{X}$'s are pairwise grouped. 
Right: The same for $M=3$ worlds. The Keldysh contour for the degrees of freedom of environment $b$ (black line) is closed encompassing all the worlds. For relevant diagrams,
$\hat{X}$ are either in the same, or different worlds: This gives two parts of ${\cal F}_M$.
}
\label{fig_worlds}
\end{figure}

To evaluate the R\'{e}nyi entropy flow of $M$-th order to an environment $b$ we need to use the perturbation theory for the $M$-th power of its density matrix, { ${\rm Tr}_b\left\{\rho_b(t)\right\}^M$}. To this end, we consider $M$ copies of the world consisting of the quantum system and the environments \cite{Nazarov}, each world bringing its own double Keldysh contour. {The contour for  the degrees of freedom of environment $b$, defining the ordering of $\hat{X}_b$, encompasses all the worlds imposing the matrix multiplication of $\rho_b$ required and finally closes (see right diagram in Fig.\ref{fig_worlds}) .} For all other degrees of freedom, the bra and ket parts of the contours are closed within each world providing the partial trace over these degrees of freedom: That yields $\rho_b$ for each world. The relevant diagrams are pairwise-grouped. For those arising from the environments other than $b$, both operators are within the same world. Summation over these diagrams reproduces evolution equation (\ref{eq:Bloch}).  The operators in diagrams from environment $b$ can be either in the same world, or in different worlds. The same-world diagrams have already been considered in Ref. \cite{Nazarov}. The different-world diagrams though contain non-diagonal elements of the system density matrix and are thus specific for the case of coherent drive and degeneracies.

In this paper, we restrict ourselves to a simple case when the transition rates induced by environment $b$ are smaller than those induced by others. The environment $b$ is thus {\it probe} one  and hardly affect the density matrix of the system. In this case, R\'{e}nyi entropy flow to the environment $b$ is directly given by the second-order diagrams encompassing two operators $\hat{X}^{(b)}$.  The diagrams are expressed in terms of the generalized correlators of two 
$\hat{X}^{(b)}$ that contain multiple powers of $\rho_b$,
\begin{equation}
 {S^{N,M}_{mn,pq}(\tau) \equiv  Tr_b \{ \hat{X}_{mn}(t)\rho_{b}^{N}\hat{X}_{pq}(t+\tau)\rho_{b}^{M-N} \} / Tr_b\{ \rho_{b}^M\} }
\end{equation}
and, for general $\rho_b$, do not correspond to any physical quantities. 
However, we derive that for the probe environment in the state of thermal equilibrium the correlators obey the generalized {Kubo-Martin-Schwinger (KMS) } \cite{{KMS}} relation  (see Appendix C),
\begin{equation}
{S^{N,M}_{mn,pq}(\nu) =  \bar{n}_B(M\nu/T) e^{\beta\nu N} \tilde{\chi}_{mn,pq}(\nu)}
\end{equation}
and therefore are all expressed in terms of the dissipative susceptibilities. In derivation, we assume that $\tilde{\chi}$ does not depend on temperature. If this is not so, $\tilde{\chi}$ is taken at $\beta^*=\beta M$. 

Collecting all diagrams (see Appendix B), we obtain for ${\cal F}_M$ the following expression:
\begin{eqnarray}
&&{\cal F}_M = \frac{M n_B(M\omega/T)}{n_B((M-1)\omega/T) n_B(\omega/T) \omega} (Q_i - Q_c) \\ \nonumber
&&   Q_i = \omega\  \left\{ \sum_{mnp;\eta_{np}=1} 
\rho_{mn} \tilde{\chi}_{pm,np}(\omega)  (1+n_B(\omega/T))  \right. \\ && \left. \ \qquad \quad  - \sum_{mnp; \eta_{pm}=1}    \rho_{m{n}} \tilde{\chi}_{np,pm}(\omega) n_B(\omega/T)\  \right\}   \\ 
&&{Q_c =  \omega  \sum_{mnpq; \eta_{pq}=1} \rho_{nm} \rho_{qp} \tilde{\chi}_{mn,pq}(\omega)}
\end{eqnarray}
The R-flow is naturally separated onto two parts, which we name  \emph{incoherent} and \emph{coherent}.
The same-world diagrams contribute to the incoherent part that is proportional to $Q_i$. $Q_i$ is linear in $\rho$ so that is an observable. The different-world diagrams form the coherent part $\propto Q_c$ that is quadratic in $\rho$ and in principle would not be observable. The $M$ dependence is identical for both parts.

Let us \emph{interpret} the parts and the quantities $Q_{i,c}$. Inspection of the rates in Eq. (\ref{eq:Bloch}) unambiguously identifies $Q_i$ with an observable: the total energy flow to the probe environment. The terms $\propto 1+n_B$ describe absorption of energy quanta $\hbar \omega$ by the environment, while those $\propto n_B$ correspond to the emission to the system. Upon taking limit $M \to 1$, the incoherent part reproduces the textbook equation for the entropy flow, ${\cal F}_S = Q_i/T_b$. We prove that for a general situation where elementary energy transfers are not restricted to $\pm \hbar\omega$, this part of the R-flows is related to full counting statistics of energy transfers and therefore can be measured {\cite{Comingsoon}}.

The interpretation of the coherent part is more involved and interesting. To proceed, let us replace in  $H_{int}$ the operators $|m\rangle\langle n|$ with classical external forces {$f_{mn}$} that are numerically equal to the elements of the system density matrix. The time-dependence of these forces 
is given by {$f_{mn} \propto \exp(-i\omega\eta_{mn})$}. These classical forces would cause energy dissipation to the probe environment that is determined from the forces and the dissipative part of  susceptibility {$\tilde{\chi}$}. This {\it fictitious energy dissipation} is precisely $Q_c$. We stress that this is not the physical dissipation occurring in the probe environment given by $Q_i  (\neq Q_c)$. However, $Q_c$ can be extracted from the measurement results: For this, one can characterize the susceptibilities involved, measure $\rho_{mn}$ (or corresponding $\langle X_{mn}(t)\rangle$) and compute $Q_c$.

Therefore we show that both parts of R-flows can be extracted from the measurement results, although in a different way: R-flows are physical. In addition, we show that the entropy flow is not directly related to { energy flow}. Rather, 
\begin{equation}
{\cal F}_S = (Q_i - Q_c)/T_b
\end{equation}
 the difference is due to quantum coherent effects in our heat engine.

Let us discuss $M$-dependence of the R-flows. In Fig. \ref{fig_plots} (left pane) we plot ${\cal F}_M/{\cal F}_S = M n_B(M\beta \omega)/(n_B((M-1)\beta \omega) n_B(\beta \omega) \beta\omega$ that conveniently depends on $\beta\omega$ only.  We see that  { for $M \gg 1$ the ratio ${\cal F}_M/{\cal F}_S = M (1-\exp({-\beta\omega}))/\beta\omega$}, that is proportional to the number of worlds involved; the same is seen  for moderate $M$. The proportionality coefficient drops down with decreasing temperature. From the other hand, at $M \to 1$ ${\cal F}_M/{\cal F}_S \approx (M-1)$ with a coefficient not depending on temperature. This sets qualitative behavior of the curves plotted in Fig. \ref{fig_plots}. The low-temperature limit of R-flows reads 
\begin{equation}
{\cal F}_M = M (Q_i - Q_c)/\omega
\end{equation}  
(this limit does not commute with $M\to 1$ since ${\cal F}_S$ diverges at low temperatures). In the absence of coherent effects, low-temperature 
R-flow is readily interpreted semiclassically \cite{Nazarov} as number of events (in our case, $\hbar\omega$ {quantum} absorptions) per second in $M$ parallel worlds. With coherencies, such simple interpretation does not work since ${\cal F}_M$ can be negative.

\section{The Simplest quantum heat engine}

Let us illustrate the behaviors of $Q_{i,c}$ for the simplest quantum heat engine possible. It has only two states, $|0\rangle$ and $|1\rangle$ coupled by coherent drive amplitude $\Omega$, with driving frequency exactly matching the energy difference $E_1-E_0=\omega$. The relevant susceptibilities are $\tilde \chi_{01,10}(\omega)\equiv  \Gamma^b$. The main environment kept at temperature $T^*$ produces the transition rates $\Gamma_{\uparrow} = \Gamma n_B(\omega/T^*)$, $\Gamma_{\downarrow} = \Gamma (1+n_B(\omega/T^*))$
while the probing environment produces similar rates $\Gamma^{b}_{\uparrow} = \Gamma^{b} n_B(\omega/T_b)$, $\Gamma^{b}_{\downarrow} = \Gamma^{b} (1+n_B(\omega/T_b))$ with $\Gamma^{b} \ll \Gamma$.

\begin{figure}[tb]
\centerline{\includegraphics[width=0.99\linewidth]{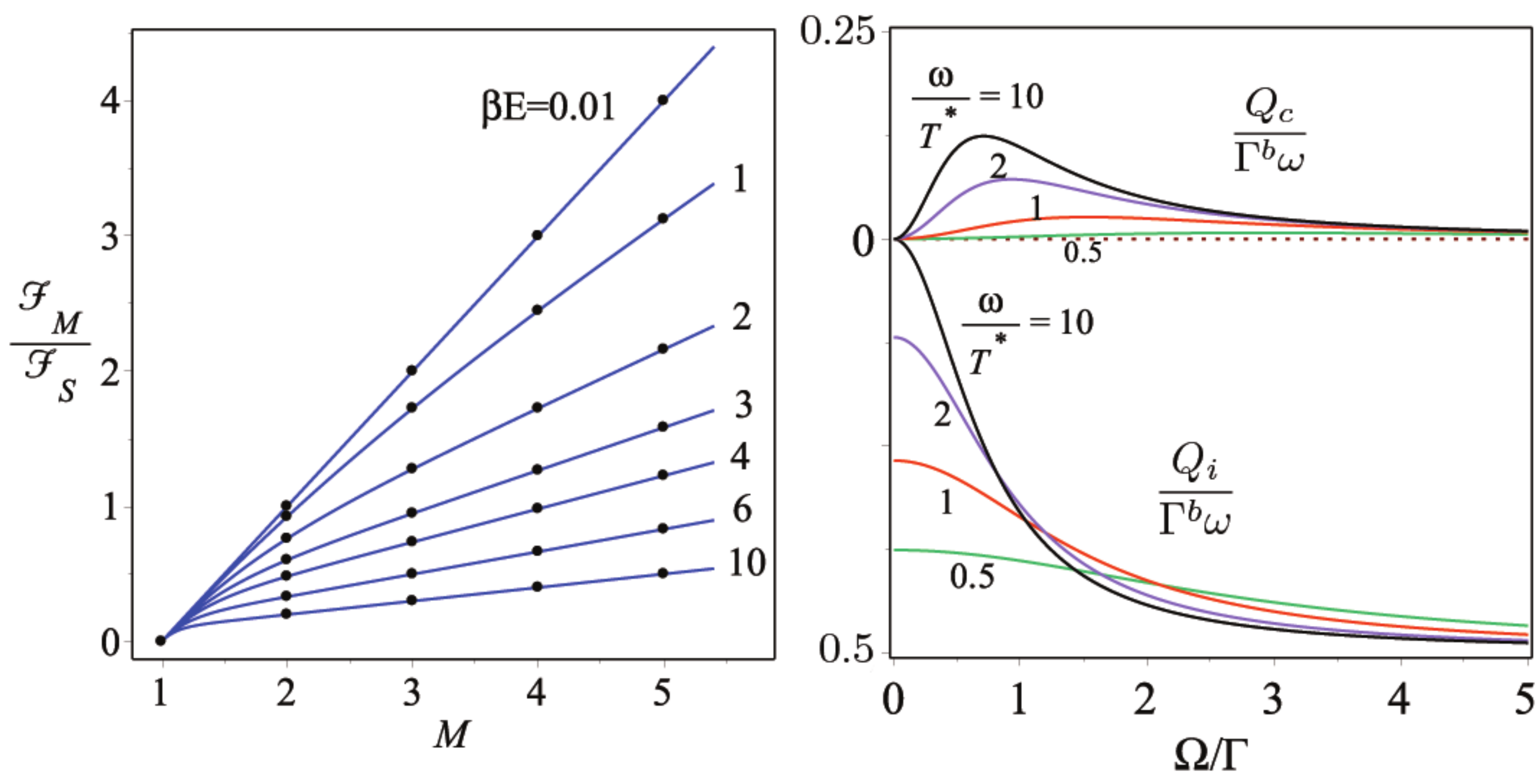}}
\caption{Left: Universal M dependence of R-flow at different temperatures,  $\beta=1/k_BT$. Flows are normalized by $(d{\cal F}/dM)_{M\to 1} \equiv \cal{F}_S$ and are suppressed at $T\to 0$.   Right: Flows $Q_c$ and $Q_i$ for the simplest QHE, at zero temperature probe environment,  for different $T^*$, versus drive strength. Coherent flow $Q_c$ reaches maximum at $\Omega \approx \Gamma$.  }
\label{fig_plots}
\end{figure}

The $Q_{i,c}$ in this case are expressed as
\begin{equation}
Q_{i}/\omega = \Gamma^{b}_{\downarrow} {p_1} -  \Gamma^{b}_{\uparrow} {p_0};\; \ \ \ 
Q_{c}/\omega = \Gamma^{b} |\rho_{01}|^2
\end{equation}
where the elements of the density matrix are determined from Eq. (\ref{eq:Bloch}) and read:   ${p_1} = 1-{p_0}$, 
\begin{eqnarray}
{{p_0} = \frac{\Gamma_{\downarrow}(\Gamma_{\downarrow}+\Gamma_{\uparrow})+\Omega^2}{(\Gamma_{\downarrow}+\Gamma_{\uparrow})^2 + 2\Omega^2},\; \ \ \ \rho_{10} = -\frac{i\Omega({p_1} -{p_0})}{\Gamma_{\downarrow}+\Gamma_{\uparrow}}} 
\end{eqnarray}  
The plots of the $Q_{i,c}$ versus drive strength are given in Fig. \ref{fig_plots} (right panel) for {\it zero} $T_b$ and different $\omega/T^*$. The coherent dissipation $Q_c$ is absent in the absence of the drive, reaches maximum $\Gamma^{b}\omega/8$ at $T^* =0$, and vanishes upon increasing $\Omega$ since non-diagonal elements of $\rho$ vanish in this limit. Finite $T^*$ suppresses the coherence and therefore {$Q_c$}. The heat flow $Q_i$ at $T^*=0$ is absent at $\Omega=0$ since the system is not excited. It increases and saturates at $\Gamma^{b}\omega/2$ for $\Omega \gg \Gamma$ when the states $|0\rangle$ and $|1\rangle$ are equally populated. At finite $T^*$, $Q_i$ is present in the absence of the drive as well. 

\section{Discussion}

Before summarizing, let us shortly outline how to compute R-flows to an environment that essentially disturbs the dynamics of the system: For the example considered this implies $\Gamma^{b} \simeq \Gamma$. In this case, the summation of the second-order diagrams leads to a linear evolution equation that generalizes Eq. (\ref{eq:Bloch}). This equation is for a matrix $R$ that is an analog of  density matrix of $M$ copies of the system and is indexed by a compound $I \equiv \{i_1, \dots, i_M \}$ encompassing all the worlds. The linear equation has a set of eigensolutions $R(t) \simeq \exp(- \Lambda t)$. In distinction from a usual equation for density matrix, there is no solution with $\Lambda=0$. The R-flow is shown to be given by ${\cal F}_M = \Lambda_0$, $\Lambda_0$ being the eigenvalue which is closest to $0$. The eigenvalues for a given number of worlds $M$ and concrete situation can be readily solved numerically. However, the analytical continuation to arbitrary $M$ is not evident for this moment and requires further research.   
    
To conclude, we have computed R\'{e}nyi entropy flows from a generic quantum heat engine to a probe environment and obtained Shanon entropy flows by taking limit $M\to1$. The flows are expressed in terms of two quantities $Q_{i,c}$ with $Q_i$ being the heat flow and $Q_c$ being an energy dissipation for the situation where the driven heat engine is replaced by fictitious coherent time-dependent classical forces. Both quantities are measurable.   
The entropy flow is proportional to $Q_i-Q_c$.  This is in contrast with frequently used \cite{Seifert} relations for entropy production along classical stochastic trajectory and implies that the concept of (R\'{e}nyi) entropy flows requires revision and clarification in quantum case.

\begin{acknowledgments}
The research leading to these results has received funding from the European Union Seventh Framework Programme INFERNOS (FP7/2007- 2013) under Grant Agreement no. 308850.
\end{acknowledgments}

\appendix

\section{Definition of R\'{e}nyi entropy and its flow}

Our definition of Renyi entropy differs by a constant factor from the widely used one $S_M=(1-M)^{-1} Tr(\rho^M)$.  Since we compute the flows, which are time-derivatives of the Renyi entropy log, i.e. ${\cal F}_M =(1/S_M) d S_M /dt$, the constant factor $(1-M)^{-1}$ in the definition cannot and does not play any role. In fact, the ``standard'' definition may lead to the confusions: it looks like the Shannon entropy can be obtained by taking the limit of Renyi entropy while in fact it is the limit of its derivative with respect to $M$. Therefore, in using our definition, one must notice that the flow of Shannon entropy is provided by ${\cal F}_S =\lim_{M \to 1}  \partial {\cal F}_M/\partial M$.

\

\section{Diagrammatic R\'{e}nyi entropy flows}

Let us compute the R-flows from expansion of Bloch equation in the second order of interaction Hamiltonian.  Considering that  far in the past the coupling between system and environment is absent, the evolution is formally:
\begin{equation}
\rho(t) = \mathcal{T} e^{ i \int_{-\infty}^t d\tau H_{int}(\tau) }\  \rho \ \tilde{\mathcal{T}} e^{  -i \int_{-\infty}^t d\tau H_{int}(\tau) }
\label{eq. app formal evol} 
\end{equation}
$T\exp$ ($\tilde{T}\exp$) refer to forward time ordering (backward time ordering).  Without loss of generality, the system-bath Hamiltonian can be taken $H_{int}=H_sH_b$, where $H_{s(b)}$ acts on the system (bath), given a Gaussian correlations of the bath: $\langle H_b(t_2) H_b(t_1)\rangle=tr_b\{H_b(t_1)H_b(t_2)\rho_b\}$ . In the second-order expansion, we place one $H_{int}$ at $t$ and the second one at any time before it, say $t-\tau$ for $0\leq\tau<\infty$. Without loss of generality we can set  the global time to $t=0$. The system density matrix in interaction picture $\rho_s=U_s(0,t)\rho'_s(t)U_s(t,0)$ evolves according to

\begin{eqnarray}\nonumber
\frac{d\rho_s}{dt}&=&  \int_0^\infty d\tau \left\{ \langle H_b(-\tau)  H_b(0) \rangle  H_s(0) \rho_s H_s(-\tau) \right.  \\ \nonumber && \qquad \quad \left.+ \langle H_b(0)  H_b(-\tau) \rangle  H_s(-\tau) \rho_s H_s(0)\right.  \\ \nonumber && \qquad \quad \left. - \langle H_b(0)  H_b(-\tau) \rangle  H_s(0) H_s(-\tau) \rho_s  \right.  \\ \nonumber&& \qquad \quad \left. + \langle H_b(-\tau)  H_b(0) \rangle  \rho_s  H_s(-\tau) H_s(0)\right\}\\
\label{eq. app flux rho}
\end{eqnarray}

Given that Renyi entropy is $S_M={\rm Tr} \rho^M$,  its flux $dS_M/dt$ can be determined directly from the generalization of Eq. (\ref{eq. app flux rho}). 
Evolution of $M$ copies of $\rho_1^M$ can influence more than one copy of the worlds. In this sense the evolution of Renyi entropy is more complex than Eq.   (\ref{eq. app flux rho}) because different worlds may exchange energies. For this aim calculating a generalized correlator  $\langle H_b(0) \rho_b^N  H_b(-\tau) \rho_b^{M-N} \rangle$ with $0 \leq N \leq M$ is required. We use the following diagrams to evaluate the partial evolution.  In the diagrams the solid (black) line denotes evolution of the system and  narrow double (white) line the rest of a world except its system.

In a typical diagrams with $M$ worlds, given that there are $N$ worlds between the operators $A(t)$ and $B(t+\tau)$, the Fourier transformed correlations consist of two parts:  $S_{A,B}^{N,M}$ and  $\Pi_{AB}^{N,M}$. These two are related through a generalized Kramers-Kronig relation. The forward correlator is
\begin{eqnarray}\nonumber
\label{eq. SPi def}
\int_{0}^{\infty}d\tau e^{i\omega\tau} {{\rm Tr}\left( A\left(0\right)\rho_{env}^{N}B\left(\pm \tau\right)\rho_{env}^{M-N}\right)/{\rm Tr} \left( \rho_{env}^{M}\right) }&& \\    \equiv \frac{1}{2}S_{AB}^{N,M}\left(\pm \omega\right) \pm i\Pi_{AB}^{N,M}\left(\pm \omega\right)\ \ \ &&
\end{eqnarray}
with the following properties: 
\begin{eqnarray}
S_{AB}^{N,M}(-\omega) &=& S_{AB}^{M-N,M}(\omega), \\
\Pi_{AB}^{N,M}\left(-\omega\right)&=&-\Pi_{BA}^{M-N,M}\left(\omega\right),\\  \Pi_{AB}^{N,M}\left(\omega\right) &=& -\frac{1}{2\pi}\frac{\int dz S_{AB}^{N,M}\left(z\right)}{z-\omega}.
\end{eqnarray}

\subsubsection{Single world diagrams:}

Different time order of the two interactions applied in one world  provides the following four diagrams for each world:

\begin{figure}[h]
\centerline{\includegraphics[width=0.5\linewidth]{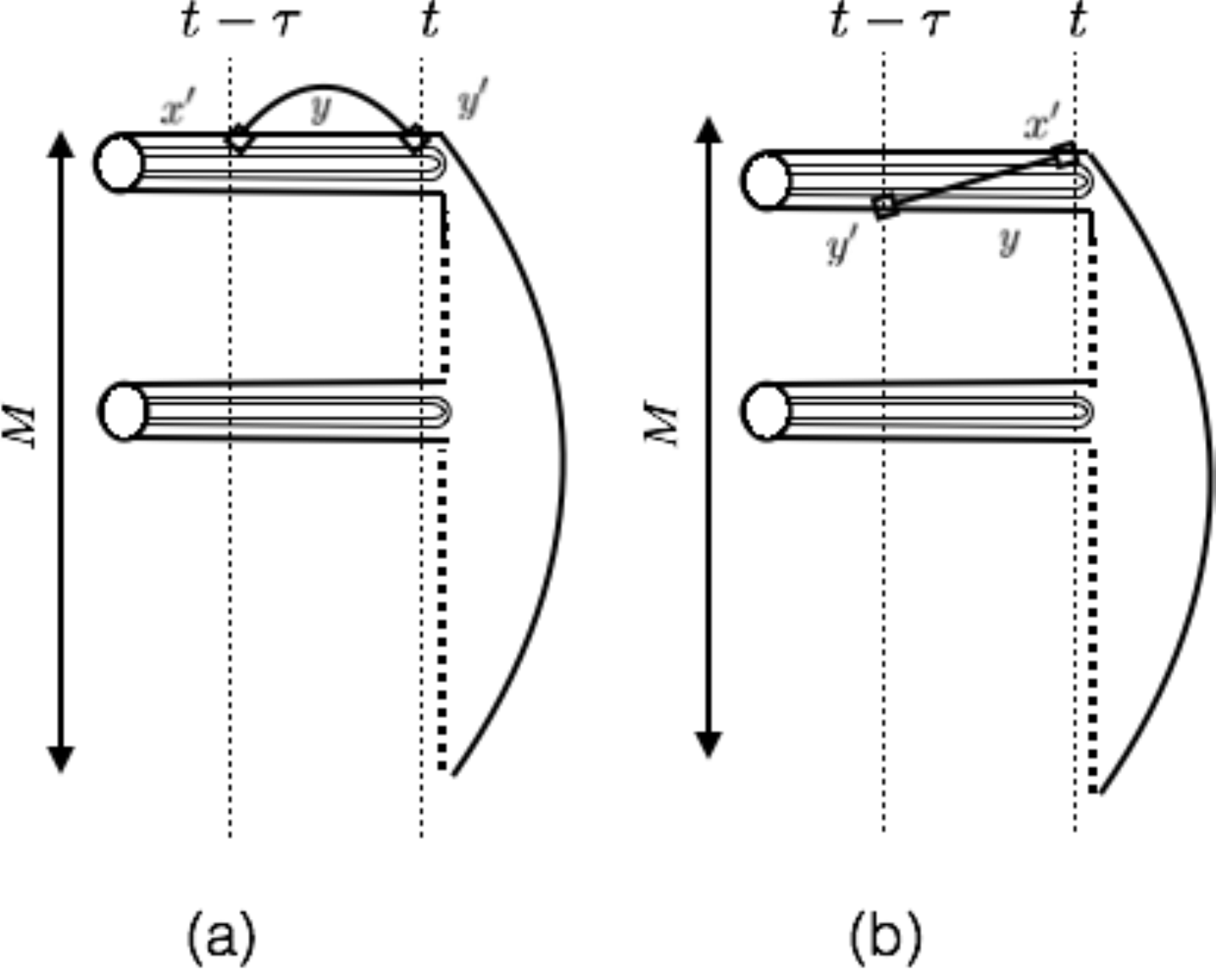}\includegraphics[width=0.5\linewidth]{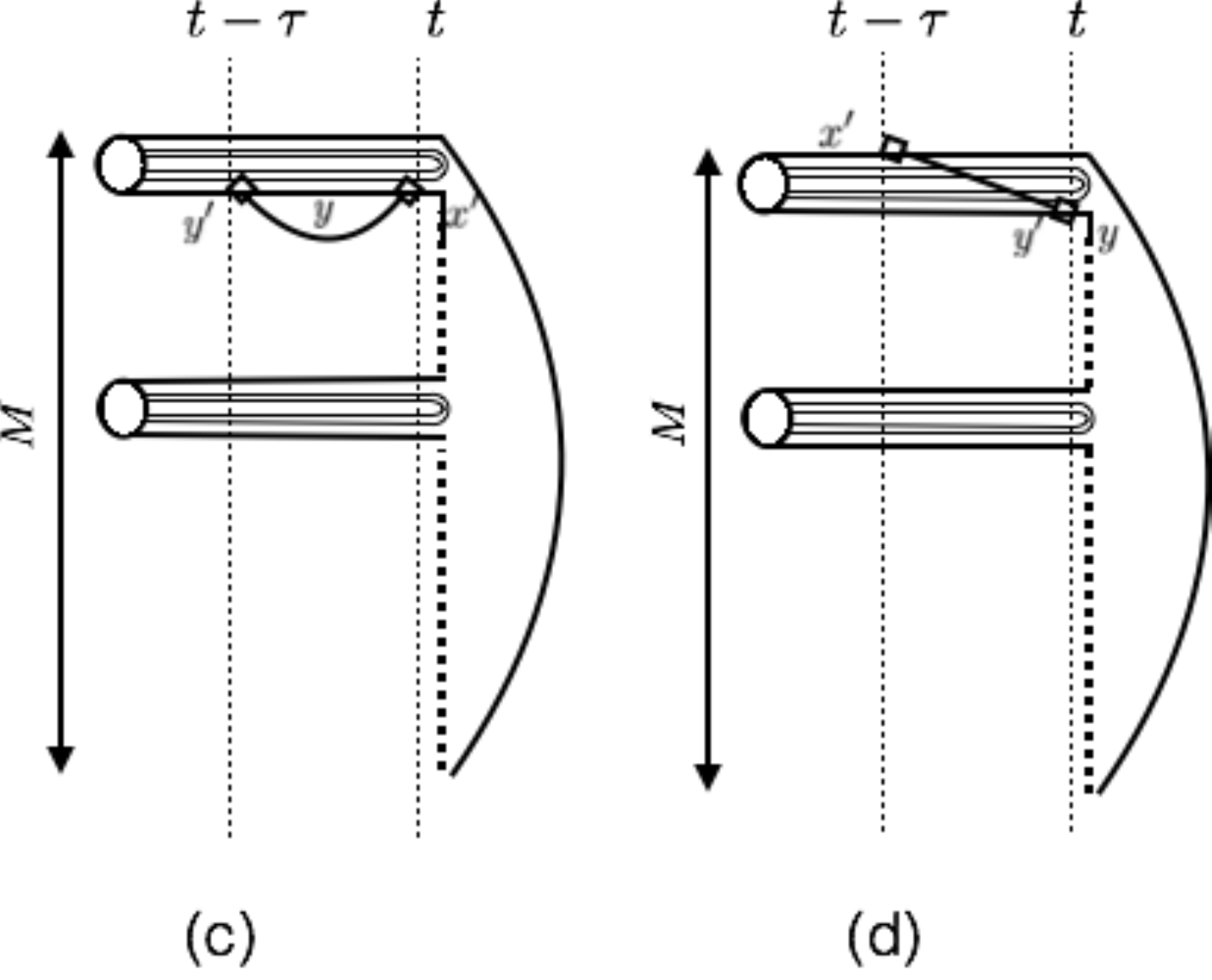}}
\end{figure}
Determining the diagrams from Eq. (\ref{eq. app flux rho}) and using the relations using Eq. (\ref{eq. SPi def}) over the diagram (a-d) can make the single world dynamics into the following form: 
\begin{eqnarray}
\nonumber
&&  \sum_{x'y'y}\rho_{x'y'} \left(-\frac{1}{2}S_{y'y,yx'}^{0,M}\left(\omega\eta_{yx'}\right)+i\Pi_{y'y,yx'}^{0,M}\left(\omega\eta_{yx'}\right)\right.\\  \nonumber & & \ \ \ \qquad \ \qquad \left. +\frac{1}{2} S_{yx',y'y}^{1,M}\left(\omega\eta_{y'y}\right)- i\Pi_{yx',y'y}^{1,M}\left(\omega\eta_{y'y}\right)\right.\\  \nonumber & & \ \ \ \qquad \ \qquad \left.  -\frac{1}{2}S_{y'y,yx'}^{0,M}\left(-\omega\eta_{y'y}\right)-i\Pi_{y'y,yx'}^{0,M}\left(-\omega\eta_{y'y}\right) \right.\\  \nonumber & & \ \ \ \qquad \ \qquad \left.+ \frac{1}{2}S_{yx',y'y}^{1,M}\left(-\omega\eta_{yx'}\right)+i\Pi_{yx',y'y}^{1,M}\left(-\omega\eta_{yx'}\right)\right)\\ 
\label{eq. app1}
\end{eqnarray}

Due to the conservation of energy in these diagrams $E_{y'}-E_{y}=E_{x'}-E_y$ which means $\eta_{yx'}=-\eta_{yy'}$. Substituting this in Eq. (\ref{eq. app1}) gives 

\begin{equation}
 \sum_{x'y'y}\rho_{x'y'} \left(-S_{yx',y'y}^{M,M}\left(\omega\eta_{y'y}\right)+ S_{yx',y'y}^{1,M}\left(\omega\eta_{y'y}\right) \right)
\label{eq. app1-2}
\end{equation}
\\

\subsubsection{Multiworld diagrams:}

Different time orders of the two interactions applied each in one world, different from that of the other one,  provide the following typical diagrams:

\begin{figure}[h]
\centerline{\includegraphics[width=0.5\linewidth]{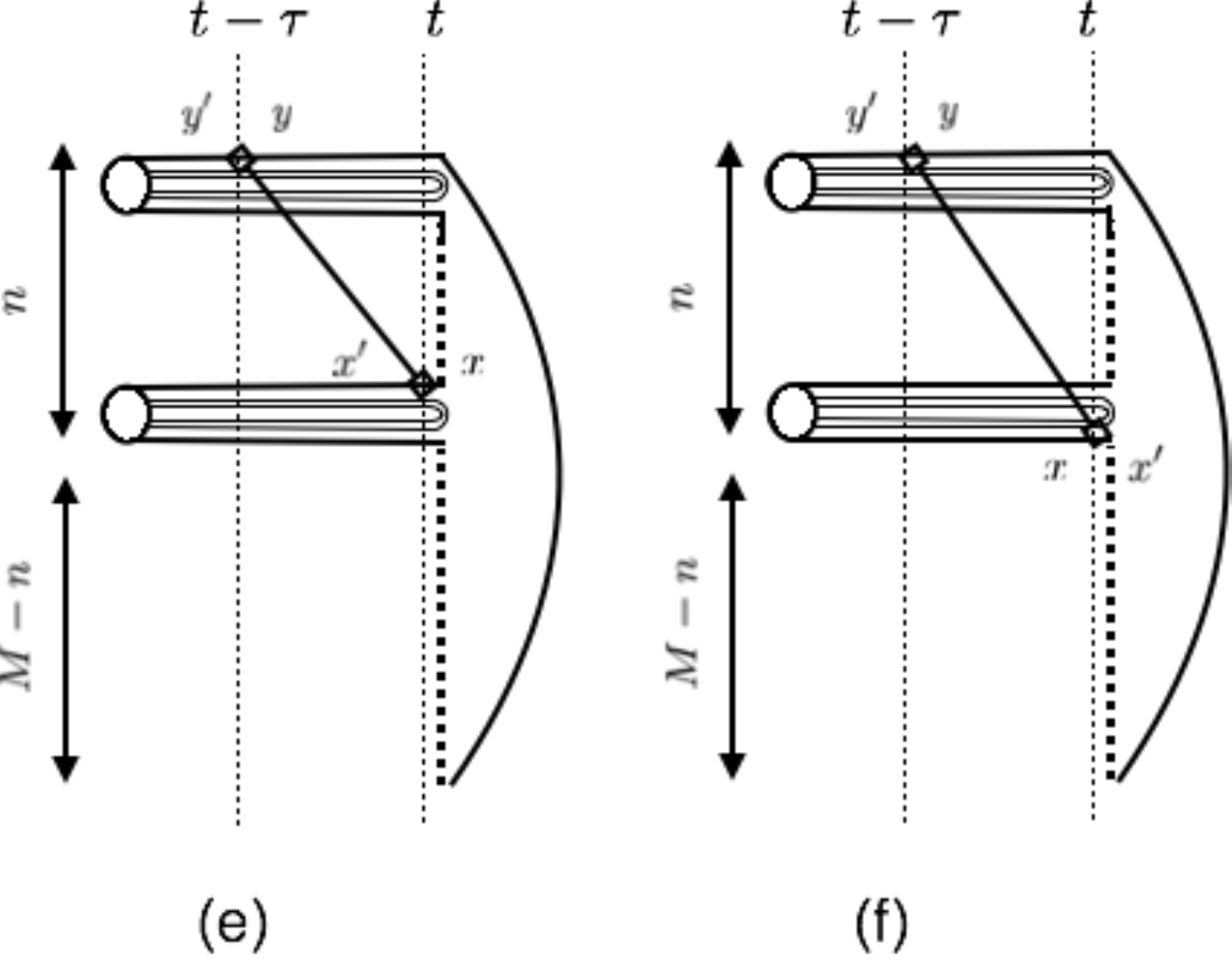}\includegraphics[width=0.5\linewidth]{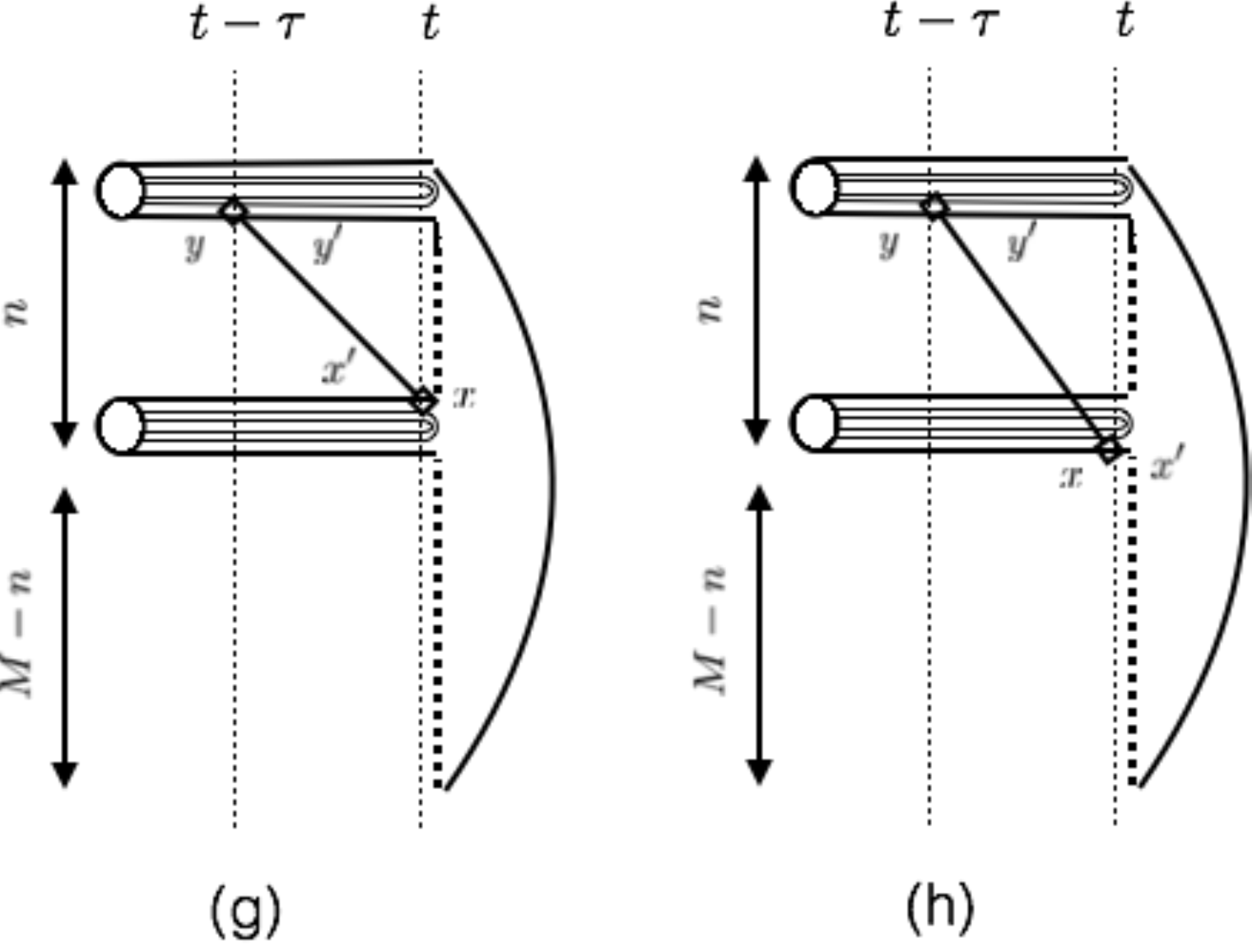}}
\centerline{\includegraphics[width=0.5\linewidth]{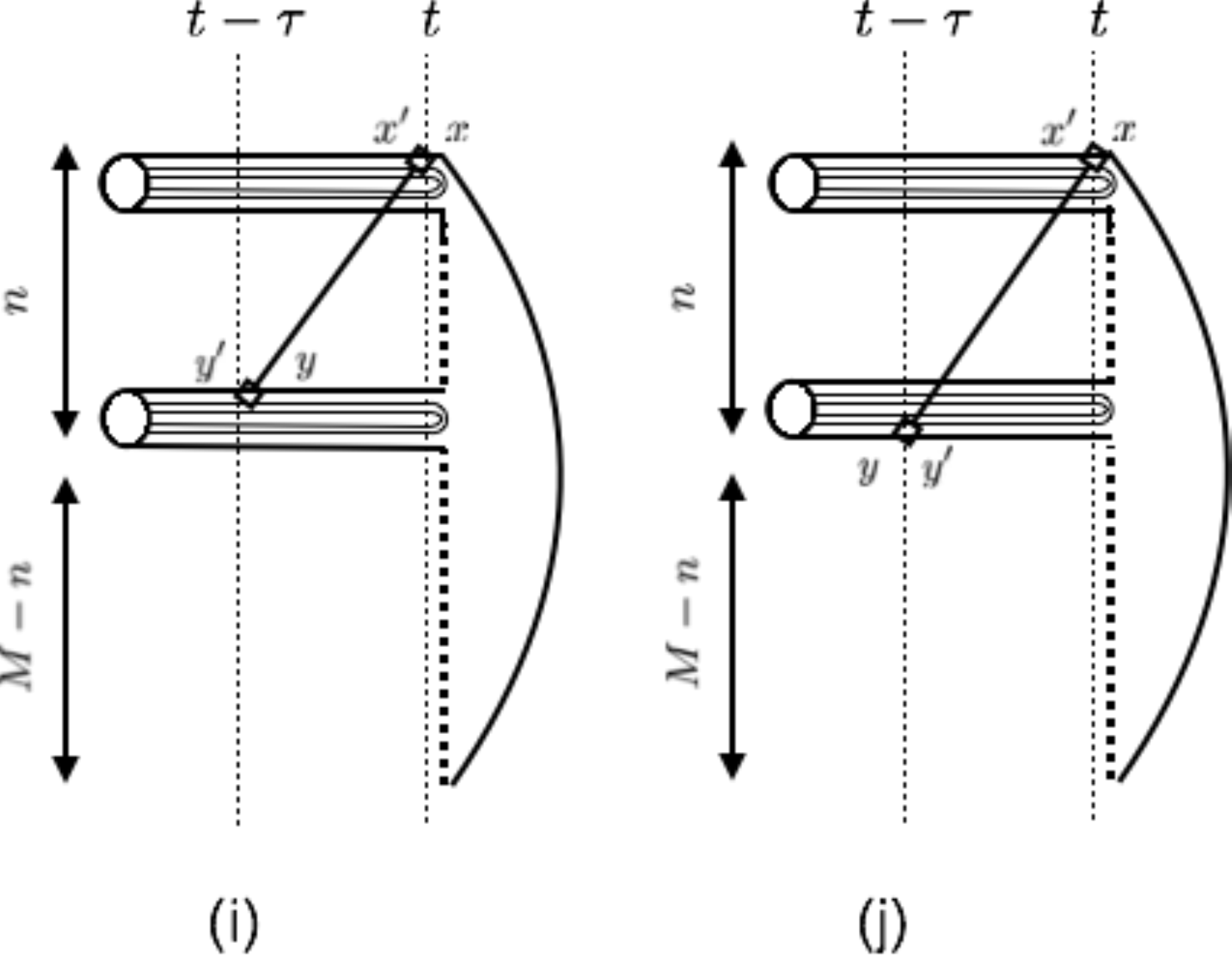}\includegraphics[width=0.5\linewidth]{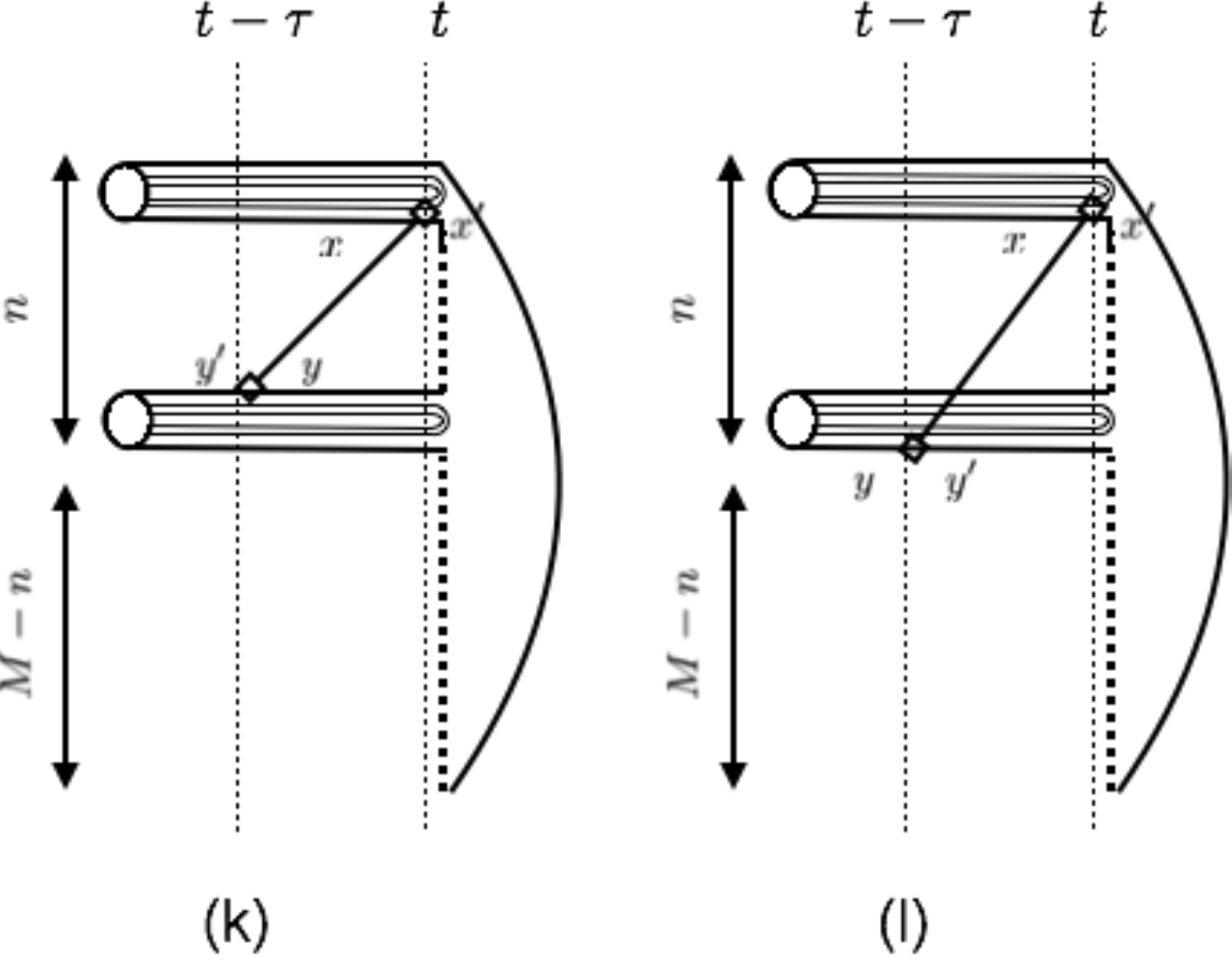}}
\end{figure}

By obtaining the diagrams from Eq. (\ref{eq. app flux rho}) and using the relations using Eq. (\ref{eq. SPi def}) one can get the sum of the forward propagating diagrams (e-h) and backward-propagating ones in diagrams (i-l):

\begin{widetext}
\begin{eqnarray}\nonumber
&&\sum_{xx'yy'}\rho_{x'x}\rho_{y'y}\left(-S_{yy',xx'}^{n-1,M}\left(-\omega\eta_{yy'}\right)-2i\Pi_{yy',xx'}^{n-1,M}\left(-\omega\eta_{yy'}\right) +\frac{1}{2}S_{yy',xx'}^{n,M}\left(-\omega\eta_{yy'}\right)+i\Pi_{yy',xx'}^{n,M}\left(-\omega\eta_{yy'}\right)\right.\\ \nonumber &&  \qquad \qquad   \left.  + \frac{1}{2}S_{yy',xx'}^{n-2,M}\left(-\omega\eta_{yy'}\right)+i\Pi_{yy',xx'}^{n-2,M}\left(-\omega\eta_{yy'}\right) \right)\\ \nonumber && 
 +\sum_{xx'yy'}\rho_{x'x}\rho_{y'y}\left(- S_{xx',yy'}^{n-1,M}\left(\omega\eta_{yy'}\right)+2i\Pi_{xx',yy'}^{n-1,M}\left(\omega\eta_{yy'}\right) +\frac{1}{2} S_{xx',yy'}^{n,M}\left(\omega\eta_{yy'}\right)-i\Pi_{xx',yy'}^{n,M}\left(\omega\eta_{yy'}\right)\right. \\   &&  \qquad \qquad \left. +\frac{1}{2}S_{xx',yy'}^{n-2,M}\left(\omega\eta_{yy'}\right)-i\Pi_{xx',yy'}^{n-2,M}\left(\omega\eta_{yy'}\right) \right),
\label{eq. app2}
\end{eqnarray}
where $n=2$ to $n=M'$. This $M'$ shows the maximum number of worlds between two interactions in our diagrams.   If the first interaction is in the topmost world, the second one can go from the second world up to the  bottommost world. This shows $M'$ is $M$. However, if we ignore the first world and put the first interaction on the second world from top, then $M'$ becomes $M-1$. This continues until $M'$ becomes 2.  The total summation of all diagrams becomes:

\begin{eqnarray}\nonumber
&& \sum_{xx'yy'}\rho_{x'x}\rho_{y'y}\sum_{M'=2}^{M}\sum_{n=1}^{M'-1}\left(- S_{xx',yy'}^{n,M}\left(\omega\eta_{yy'}\right)+ \frac{1}{2}S_{xx',yy'}^{n+1,M}\left(\omega\eta_{yy'}\right) + \frac{1}{2}S_{xx',yy'}^{n-1,M}\left(\omega\eta_{yy'}\right)\right. \\ && \nonumber \qquad \qquad \qquad \ \ \ \ \left.  +2i\Pi_{xx',yy'}^{n,M}\left(\omega\eta_{yy'}\right) -i\Pi_{xx',yy'}^{n-1,M}\left(\omega\eta_{yy'}\right) -i\Pi_{xx',yy'}^{n+1,M}\left(\omega\eta_{yy'}\right)\right) \\ \nonumber
&=&  \frac{M}{2} \sum_{xx'yy'}\rho_{x'x}\rho_{y'y}\left(S_{xx',yy'}^{0,M}\left(\omega\eta_{yy'}\right)- S_{xx',yy'}^{1,M}\left(\omega\eta_{yy'}\right) -S_{xx',yy'}^{M-1,M}\left(\omega\eta_{yy'}\right)+S_{xx',yy'}^{M,M}\left(\omega\eta_{yy'}\right)\right. \\ &&  \qquad \qquad \qquad \ \ \ \ \left.  -2i\Pi_{xx',yy'}^{0,M}\left(\omega\eta_{yy'}\right) +2i\Pi_{xx',yy'}^{1,M}\left(\omega\eta_{yy'}\right) +2i\Pi_{xx',yy'}^{M-1,M}\left(\omega\eta_{yy'}\right) -2i\Pi_{xx',yy'}^{M,M}\left(\omega\eta_{yy'}\right)\right) 
\label{eq. app3}
\end{eqnarray}

 Let us look at a typical $\Pi$-term: $\sum_{xx',yy'}\rho_{x'x}\rho_{y'y} \Pi^{a,M}_{xx',yy'}(\omega \eta_{yy'})$ with $a=0, 1, M-1, M$. In the energy eigenbasis of four states $n,m,k,l$ with the property $E_n-E_m=E_l-E_k>0$  the series summation is expanded  into  $\rho_{nm}\rho_{kl} [\Pi^{a,M}_{mn,lk}(\omega )- \Pi^{M-a,M}_{mn,lk}(\omega )]+\rho_{mn}\rho_{lk} [-\Pi^{M-a,M}_{kl,nm}(\omega )+ \Pi^{a,M}_{kl,nm}(\omega)] $. By substituting in Eq. (\ref{eq. app3})   all $\Pi$ terms vanish.
 
 As a result the Renyi entropy flow becomes

 \begin{eqnarray}\nonumber
\frac{1}{S_{M}}\frac{dS_{M}}{dt} & = & M \left( \sum_{x'y'y}\rho_{x'y'} \left(-S_{yx',y'y}^{M,M}\left(\omega\eta_{y'y}\right)+ S_{yx',y'y}^{1,M}\left(\omega\eta_{y'y}\right) \right) \right. \\
 && \left. \ \ \  +\frac{1}{2}\sum_{xx'yy'}\rho_{x'x}\rho_{y'y}\left(S_{xx',yy'}^{0,M}\left(\omega\eta_{yy'}\right)- S_{xx',yy'}^{1,M}\left(\omega\eta_{yy'}\right) -S_{xx',yy'}^{M-1,M}\left(\omega\eta_{yy'}\right)+S_{xx',yy'}^{M,M}\left(\omega\eta_{yy'}\right)\right) \right)
\label{eq. app5}
\end{eqnarray}
\end{widetext}

\section{Generalized KMS}
The generalized correlator of two operators $A$ and $B$ is defined [see Eq. (4)] as 
\begin{equation*}
S^{N,M}_{AB}\left(\omega\right)  =  \int d \tau e^{i\nu \tau} {\rm Tr} \{ A(0) \rho_b^N B(\tau) \rho_b^{M-N} \} / {\rm Tr}{\rho_b^M}
\end{equation*}
 
 This correlator in the energy eigenbasis can be rewritten in matrix form

\begin{eqnarray}\nonumber
&& S_{nm,mn}^{N,M}=   \int d \tau e^{i\nu \tau} \\ \nonumber &&   \left( A_{nm} \frac{e^{-\beta N E_m}}{Z(\beta)^N} B_{mn} e^{i (E_m-E_n)\tau} \frac{e^{-\beta E_n (M-N)}}{Z(\beta)^{M-N}} \right) \frac{Z(\beta)^M}{Z(\beta M)} \\  
&&= 2\pi  \delta\left( E_m-E_n+\nu \right) \frac{A_{nm}  B_{mn} e^{-\beta E_n M} }{Z(\beta M)}e^{\beta N \nu}
\label{eq. app kms f}
\end{eqnarray}
where $Z(\beta)$ is the partition function defined as $Z(\beta)=\sum_i e^{-\beta E_i}$. The standard correlator is  $S_{AB}\left(\omega\right)  =  \int d \tau \exp({i\nu \tau}) {\rm Tr} \{ A(0)  B(\tau) \rho_b \}/ {\rm Tr}{\rho_b}$ which after simplification becomes equal $ 2\pi  \delta\left( E_m-E_n+\nu \right) A_{nm}  B_{mn} e^{-\beta E_n } /Z(\beta )$, where the KMS relation links this to dynamical susceptibility: $S_{AB}(\nu)=\tilde{\chi}_{AB}(\nu)n_B(\nu/T)$. By substituting this in Eq. (\ref{eq. app kms f}) a generalized KMS relation is obtained:
\begin{equation}
S^{N,M}_{AB}\left(\omega\right) =  n_B\left(M\omega/T\right) e^{\beta\omega N} \tilde{\chi}_{AB}\left(\omega\right)\label{eq. app kms1}
\end{equation}
\\

\section{Renyi entropy flow}

By substituting the generalized KMS relation (\ref{eq. app kms1}) in Eq. (\ref{eq. app5}) the Renyi entropy flow is determined based on susceptibility:
\begin{widetext}
 \begin{eqnarray}\nonumber
\frac{1}{S_{M}}\frac{dS_{M}}{dt}   &=& - M\sum_{x'y'y}\rho_{x'y'}  \tilde{\chi}_{yx',y'y}\left(\omega\eta_{y'y}\right)   \frac{n_B \left(M\omega\eta_{y'y}/T\right)}{n_B\big(\left(M-1\right)\omega\eta_{y'y}/T\big)} e^{\beta  \omega\eta_{y'y}}    \\ \nonumber
  &  & + \frac{M}{2}\sum_{xx'yy'}\rho_{x'x}\rho_{y'y}  \tilde{\chi}_{xx',yy'}\left(\omega\eta_{yy'}\right)  \frac{n_B(M\omega\eta_{yy'}/T)}{n_B\big( (M-1) \omega\eta_{yy'}/T\big)}\left( e^{\beta \omega\eta_{yy'}} -1\right) \\
\label{eq. app6}
\end{eqnarray}
\end{widetext}

\end{document}